\documentclass[12pt]{article}  
\usepackage{latexsym} 
\usepackage{amssymb,amsmath} 
\numberwithin{equation}{section}
\newcommand{\mL}{{\mathcal L}}
\title{Modified Hamiltonian formalism for higher-derivative theories}
\author{
K. Andrzejewski,  J. Gonera, P. Machalski, P. Ma\'slanka\footnote{e-mail: pmaslan@uni.lodz.pl}
\vspace*{1cm}
\\
Department of Theoretical Physics II, University of {\L}\'od\'z,\\
Pomorska 149/153, 90 - 236 {\L}\'od\'z, Poland.}
\date{}
\begin{document}
\maketitle
\begin{abstract}
The alternative version of Hamiltonian formalism for higher-derivative theories is proposed. As compared
with the standard Ostrogradski approach it has the following  advantages:
(i) the Lagrangian, when expressed
in terms of new variables yields proper equations of motion; no additional Lagrange multipliers are necessary 
(ii)  the Legendre transformation  can be performed in a straightforward way provided the Lagrangian is nonsingular in
Ostrogradski sense.
The generalizations to singular Lagrangians as well as field theory are presented.
\end{abstract}
\section*{Introduction \label{sec:1}}
It is a long-standing problem whether and why it is sufficient to use in physics the Lagrangians containing 
only first order time derivative. This is the more intriguing that adding higher derivatives may improve
our models in some respects, like ultraviolet behaviour \cite{1b,2b} (in particular, making modified
gravity renormalizable \cite{3b} or even asymptotically free \cite{4b}); also, higher-derivative Lagrangians appear to be a useful tool to describe some interesting models like relativistic particles with rigidity, curvature and torsion \cite{19b}  Moreover, almost any effective theory
obtained by integrating  out some degrees of freedom (usually, but not always, those related to high energy
excitations) of the underlying "microscopical"  theory contains higher derivatives. One can argue that the 
effective theory, being an approximation to perfectly consistent quantum theory need not to be considered
and quantized separately. However, we are never sure if our  theory is the basic or effective one; therefore,
it is important to know  whether it is at all possible to quantize  the effective theory in a way which would
correctly  reproduce some aspects  of the microscopic one.
\par
First step toward the quantum theory is to put its classical counterpart in Hamiltonian form. Standard 
framework  for dealing with  higher-derivative theories on Hamiltonian level is provided by Ostrogradski 
formalism \cite{5b}-\cite{9b}.  The main  disadvantage  of the latter is that the Hamiltonian, being linear function of some 
momenta, is necessarily  unbounded from below. In general, this cannot  be cured by trying  to devise an
alternative  canonical formalism. In fact, any Hamiltonian is an integral of motion while it is by far 
not obvious that a generic  system described by higher-derivative Lagrangians  posses  globally defined 
integrals of motion, except  the one related to time translation invariance. Moreover, the instability of
Ostrogradski Hamiltonian is not related to finite domains in phase space which implies that it survives 
standard quantization procedure (i.e. cannot be cured by uncertainty principle).
\par
Ostrogradski approach has also some other disadvantages. There is no straightforward transition from the
Lagrangian to the Hamiltonian formalism. In fact, Ostrogradski approach is based on the idea that the
consecutive time derivatives of initial coordinate(s) form new coordinates $q_i\sim  q^{(i-1)}$. It appears
then that the Lagrangian cannot be viewed as a function on the tangent bundle to coordinate manifold
because it leads to incorrect equations of motion. Also, the Legendre transformation to the cotangent 
bundle  (phase space) cannot be performed. One deals with  this problem by adding Lagrange
multipliers  enforcing the proper relation between new  coordinates  and time derivatives of the original 
ones. This results in further enlarging of coordinate manifold; moreover, the theory becomes constrained
(in spite of the fact that the initial theory  may be nonsingular in the Ostrogradski sense, c.f. eq. (\ref{e:0}) below) and 
the Hamiltonian formalism  is obtained by applying Dirac constraint theory, i.e., by  reduction of the
cotangent  bundle to submanifold  endowed with sympletic structure defined by Dirac brackets.
\par
In the present paper an alternative approach is proposed. It leads directly to the Lagrangians which, being
a function on the tangent manifold, gives correct equations of motion; no new coordinate variables
need to be added. Furthermore, for Lagrangians nonsingular in Ostrogradski sense the
Legendre transformation  takes the standard form. Our approach is also applicable to the most interesting case of singular
Lagrangians (for example, those defining $f(R)$ gravities \cite{10b}).
\par
The paper is organized as follows. In Section \ref{sec:2}  we consider nonsingular Lagrangians containing second
and third order time derivatives. Constrained theories  are discussed  in Section \ref{sec:3}. The general 
formalism is applied to  mini-superspace formulation of $f(R)$ gravity \cite{11b} in Section \ref{sec:4}. In Section \ref{sec:5},
 the modifications necessary to cover the field-theoretic case
are given. In Appendix  we describe (for one degree  of freedom) the generalization of our formalism
to Lagrangians containing arbitrary high derivatives.
\section{Nonsingular Lagrangians  of second and  third order \label{sec:2}}
In this section we consider the Lagrangians  containing second and third time derivatives which are 
nonsingular in Ostrogradski sense. Ostrogradski approach is based on the idea that the consecutive 
time derivatives of the initial coordinate form a new coordinates, $q_i\sim q^{(i-1)}$. However, it 
has been suggested \cite{12b}-\cite{15b}, \cite{18b} that one can use every second derivative as a new variable, $q_i\sim q^{(2i-2)}$.
We generalize this idea by introducing new coordinates as some functions of the initial ones and their time derivatives.
Our paper is inspired by the results obtainded in Ref. \cite{13b}.
\subsection{The case of second derivatives\label{subsec:1}}
Let us start with  Lagrangians containing  time derivatives up to the second order,
\begin{equation}
\label{l35}
L=L(q,\dot q,\overset{..}{q});
\end{equation}
here $q=(q^\mu)$, $\mu=1,\ldots,N$ denotes the set of generalized coordinates. The nonsingularity 
condition of Ostrogradski reads
\begin{equation}
\label{e:0}
\det\left( \frac{\partial^2 L}{\partial \overset{..}{q^{\mu}}\partial \overset{..}{q^{\nu}}}\right)\neq 0.
\end{equation}
In order to put our theory in the first-order form we define new coordinates $q_1^\mu,q_2^\nu$:
\begin{equation}
q^{\mu}=q_1^{\mu},\quad \dot q^{\mu}=\dot q_1^{\mu},\quad \overset{..}{q}^{\mu}=\chi^{\mu}(q_1,\dot q_1,q_2),
\end{equation}
where $\chi^\mu$ are the functions specified below.
\par
We select an arbitrary function 
\begin{equation}
F=F(q_1,\dot q_1,q_2),
\end{equation}
subjected to the single condition
\begin{equation}
\label{l36}
\det\left(\frac{\partial^2 F}{\partial\dot q_1^{\mu}\partial q_2^{\nu}}\right)\neq 0.
\end{equation}
Now, $\chi^{\mu}$ are defined as a unique (at least locally due to (\ref{e:0})) solution to the following
set  of equations
\begin{equation}
\label{l40}
\frac{\partial L(q_1,\dot q_1,\chi)}{\partial \chi^{\mu}}+\frac{\partial F(q_1,\dot q_1,q_2)}{\partial \dot q_1^{\mu}}=0.
\end{equation}
The new Lagrangian, which is now a standard Lagrangian of first order, is given by 
\begin{eqnarray}
\label{l40b}
\mL(q_1,\dot q_1,q_2,\dot q_2)&=&L(q_1,\dot q_1,\chi (q_1,\dot q_1,q_2))+\frac{\partial F(q_1,\dot q_1,q_2)}{\partial q_1^{\mu}}\dot q_1^{\mu}\nonumber\\
& +&\frac{\partial F(q_1,\dot q_1,q_2)}{\partial q_2^{\mu}}\dot q_2^{\mu}+\frac{\partial F(q_1,\dot q_1,q_2)}{\partial\dot q_1^{\mu}}\chi^{\mu}(q_1,\dot q_1,q_2).
\end{eqnarray}
It differs from the initial one by an expression which becomes "on-shell" a total time derivative.
\par
The equation of motion for $q_2^\mu$ yield
\begin{equation}
\frac{\partial^2 F}{\partial\dot q_1^{\nu}\partial q_2^{\mu}}(\chi^{\nu}-{\overset{..}{q}}_1^{\nu})=0,
\end{equation}
which, by virtue of (\ref{l36}), implies
\begin{equation}
\label{l37}
\overset{..}{q}^{\mu}=\chi^{\mu}(q_1,\dot q_1,q_2).
\end{equation}
For the remaining variables  $q_1^{\mu}$ one obtains
\begin{equation}
\frac{\partial L}{\partial q_1^{\mu}}-\frac{d}{dt}\left(\frac{\partial L}{\partial \dot q_1^{\mu}}\right)+\frac{d^2}{dt^2}\left(\frac{\partial L}{\partial \chi^{\mu}}\right)=0,
\end{equation}
and taking into account  (\ref{l37})  one gets the initial Euler-Lagrange equations.
\par
It is worth to notice that, contrary to the original Ostrogradski approach, the formalism presented above
leads directly to the standard picture of Lagrangian as a function defined on the tangent bundle to
coordinate space (with no need of enlarging of the latter by adding the appropriate Lagrange multipliers).
\par
Our Lagrangian (\ref{l40b}) is nonsingular in the usual sense so one can directly pass to the Hamiltonian 
picture by performing Legendre transformation leading to canonical dynamics on cotangent bundle.
\par
To this end we define the canonical momenta
\begin{equation}
\label{l39}
p_{1\mu}\equiv \frac{\partial \mL}{\partial\dot q_1^{\mu}}=\frac{\partial L}{\partial \dot q_1^{\mu}}+
\frac{\partial^2 F}{\partial q_1^{\nu}\partial \dot q_1^{\mu}}\dot q_1^{\nu}
+\frac{\partial^2 F}{\partial\dot q_1^{\mu}\partial \dot q_1^{\nu}} \chi^{\nu}+\frac{\partial^2 F}{\partial\dot q_1^{\mu}\partial  q_2^{\nu}}\dot q_2^{\nu}+
\frac{\partial F}{\partial q_1^{\mu}},
\end{equation}
\begin{equation}
\label{l38}
p_{2\mu}\equiv\frac{\partial \mL}{\partial\dot q_2^{\mu}}=\frac{\partial F(q_1,\dot q_1,q_2)}{\partial q_2^{\mu}}.
\end{equation}
By virtue of  (\ref{l36}) the second set of equations can be uniquely solved (at least locally) for  $\dot q_1^{\mu}$
\begin{equation}
\dot q_1^{\mu}=\dot q_1^{\mu}(q_1,q_2,p_2).
\end{equation}
As for the first set  (\ref{l39}),  we note that  $\dot q_2^{\mu}$ appears (linearly) only in the fourth term 
on the RHS. Again, the same condition (\ref{l36})  allows us to solve (\ref{l39}) for  $\dot q_2^{\mu}$,
\begin{equation}
\dot q_2^\mu=q_2^\mu(q_1,q_2,p_1,p_2).
\end{equation}
\par
The Hamiltonian  $H$ is computed in standard way and the final result reads
\begin{equation}
\label{l41}
H=p_{1\mu}\dot q_1^{\mu}-L-\frac{\partial F}{\partial q_1^{\mu}}\dot q_1^{\mu}-\frac{\partial F}{\partial \dot q_1^{\mu}}\chi^{\mu},
\end{equation}
where everything is expressed in terms of  $q_1,q_2,p_1$ and $p_2$. We have checked, by direct calculation,
that  the canonical equations following from $H$ are equivalent to the initial Lagrangians ones.
\par
There exists canonical transformation which relates our Hamiltonian to the Ostrogradski one.
It reads
\begin{equation}\label{eq116}
\begin{array}{l}
\tilde q_1^{\mu} =q_1^{\mu},\\
\tilde q_2^{\mu}=f^{\mu}(q_1,q_2,p_2),\\
\tilde p_{1\mu}=p_{1\mu}-\frac{\partial F}{\partial q_1^{\mu}}(q_1,f,q_2),\\
\tilde p_{2\mu}=-\frac{\partial F}{\partial f^{\mu}}(q_1,f,q_2),
\end{array}
\end{equation}
where tilde refers to Ostrogradski variables and  $f^{\mu}(q_1,q_2,p_2)$  solve eqs.  (\ref{l38}), i.e.
$f^{\mu}=\dot q_1^{\mu}(q_1,q_2,p_2)$. The corresponding generating function $\Phi  $ has the form
\begin{equation}
\Phi(q_1,\tilde p_1,q_2,\tilde q_2)=q_1^\mu\tilde p_{1\mu}+F(q_1,\tilde q_2,q_2).\label{eq117}
\end{equation}
However, it should be stressed that Ostrogradski Hamiltonian is singular in the sense that the inverse 
Legendre transformation cannot be performed  (contrary to our case). This means that the structure of 
sympletic manifold (phase space) as a cotangent  bundle to coordinate manifold is not transparent if
Ostrogradski variables are used.
\par
Let us conclude this part with a very simple example. The Lagrangian
\begin{equation}
\label{l47b}
L=\lambda \epsilon _{\mu\nu}\dot q^{\mu}{\overset{..}{q}}^{\nu}+\frac{\beta}{2}({\overset{..}{q}}^{\mu})^2,\quad\beta\neq 0,
\quad \mu,\nu=1,2
\end{equation}
is nonsingular in Ostrogradski sense provided $\beta\neq 0$. We take
\begin{equation}
F=\alpha \dot q_1^{\mu} q_2^{\mu},\quad \alpha\neq 0.
\end{equation}
Then
\begin{equation}
\chi^{\mu}=\frac{\lambda}{\beta}\epsilon_{\mu\nu}\dot q_1^{\nu}-\frac{\alpha}{\beta}q_2^{\mu},
\end{equation}
and
\begin{equation}
\mL=-\frac{\alpha^2}{2\beta}(q_2^{\mu})^2-\frac{\lambda^2}{2\beta}(\dot q_1^{\mu})^2-
\frac{\alpha\lambda}{\beta}\epsilon_{\mu\nu}\dot q_1^{\mu}q_2^{\nu}+
\alpha\dot q_1^{\mu}\dot q_2^{\mu}.
\end{equation}
Finally, the Hamiltonian reads 
\begin{equation}
\label{l48b}
H=\frac{1}{\alpha}p_{1\mu}p_{2\mu}+\frac{\lambda^2}{2\alpha^2\beta}(p_{2\mu})^2+
\frac{\lambda}{\beta}\epsilon_{\mu\nu}p_{2\mu}q_2^{\nu}+\frac{\alpha^2}{2\beta}(q_2^{\mu})^2.
\end{equation}
It depends on an arbitrary parameter $\alpha$. One can pose the question whether any relevant physical quantity may depend on $\alpha$.
The answer is no: all physical quantities are $\alpha$-independent. Formally this can be shown using eqs~\eqref{eq116} and \eqref{eq117}.
Indeed, the function generating the canonical transformation to Ostrogradski variables reads
\begin{equation}\label{eq123}
\Phi(q_1^\mu,\tilde{p}_{1\mu},q_2^\mu,\tilde{q}_2^\mu)=
q_1^\mu\tilde{p}_{1\mu}+\alpha\tilde{q}_2^\mu q_2^\mu.
\end{equation}
The corresponding canonical transformation takes the form
\begin{equation}
\begin{array}{l}
p_{1\mu}=\tilde{p}_{1\mu},\\
q_1^\mu=\tilde{q}_1^\mu,\\
q_2^\mu=-\frac{1}{\alpha}\tilde{p}_{2\mu},\\
p_{2\mu}=\alpha\tilde{q}_2^\mu;
\end{array}
\end{equation}
when inserted into the Hamiltonian~\eqref{l48b} it yields the standard Ostrogradski Hamiltonian
\begin{equation}
\label{eq125}
H=\tilde{p}_{1\mu}\tilde{q}_2^\mu+\frac{1}{2\beta}(\tilde{p}_{2\mu})^2-\frac{\lambda}{\beta}\epsilon_{\mu\nu}\tilde{q}_2^\mu\tilde{p}_{2\nu}
+\frac{\lambda^2}{2\beta}(\tilde{q}_2^\mu)^2.
\end{equation}
It does not depend on $\alpha$. Therefore, the energy (energy spectrum in quantum theory) does not depend on $\alpha$.The role of our 
$\alpha$-dependent modification is to provide the formalism which yields standard Lagrangian dynamics and regular Legendre transformation.

The above explanation is slightly formal. We shall now look at the problem of $\alpha$ dependence from a slightly different point of view.
Let us note that the classical state our system is uniquely determined once the values of $q(t)$, $\overset{.}{q}(t)$, $\overset{..}{q}(t)$,
$\overset{...}{q}(t)$ at some moment $t$ are given. Moreover, most physically relevant quantities are constructed via Noether procedure 
(they are either conserved or partially conserved, i.e., their time derivatives are defined by transformation properties of symmetry breaking 
terms in the action). As such they are expressible in terms of $q$, $\overset{.}{q}$, $\overset{..}{q}$ and $\overset{...}{q}$.
Therefore the latter are the basic variables. One can find their quantum counterparts provided we compute the relevant Poisson brackets.

To this end we write out the canonical equations of motion following from eq.~\eqref{l48b}: 
\begin{equation}
\label{eq126}
\begin{array}{l}
\overset{.}{q}_1^\mu=\frac{1}{\alpha}p_{2\mu},\\
\overset{.}{q}_2^\mu=\frac{1}{\alpha}p_{1\mu}+\frac{\lambda^2}{\alpha^2\beta}p_{2\mu}+\frac{\lambda}{\beta}\epsilon_{\mu\nu}q_2^\nu,\\
\overset{.}{p}_{1\mu}=0,\\
\overset{.}{p}_{2\mu}=\frac{\lambda}{\beta}\epsilon_{\mu\nu}p_{2\nu}-\frac{\alpha^2}{\beta}q_2^\mu.
\end{array}
\end{equation}
They lead to the following relations
\begin{equation}
\label{eq127}
\begin{array}{l}
q^\mu=q_1^\mu,\\
\overset{.}{q}^\mu=\frac{1}{\alpha}p_{2\mu},\\
\overset{..}{q}^\mu=\frac{\lambda}{\alpha\beta}\epsilon_{\mu\nu}p_{2\nu}-\frac{\alpha}{\beta}q_2^\mu,\\
\overset{...}{q}^\mu=-\frac{2\lambda^2}{\alpha\beta^2}p_{2\mu}-\frac{2\lambda\alpha}{\beta^2}\epsilon_{\mu\nu}q_2^\nu-\frac{1}{\beta}p_{1\mu}.
\end{array}
\end{equation}
One can now find the Poisson brackets among $q$, $\overset{.}{q}$, $\overset{..}{q}$ and $\overset{...}{q}$. The nonvanishing ones read
\begin{equation}
\label{eq128}
\begin{array}{lll}
\{q^\mu,\overset{...}{q}^\nu\}=-\frac{1}{\beta}\delta_{\mu\nu},&\quad &
\{\overset{.}{q}^\mu,\overset{..}{q}^\nu\}=\delta_{\mu\nu},\\
\{\overset{.}{q}^\mu,\overset{...}{q}^\nu\}=-\frac{2\lambda}{\beta^2}\epsilon_{\mu\nu},&\quad &
\{\overset{..}{q}^\mu,\overset{..}{q}^\nu\}=\frac{2\lambda}{\beta}\epsilon_{\mu\nu},\\
\{\overset{.}{q}^\mu,\overset{..}{q}^\nu\}=\frac{4\lambda^2}{\beta^3}\delta_{\mu\nu}, &\quad &
\{\overset{.}{q}^\mu,\overset{..}{q}^\nu\}=\frac{8\lambda^3}{\beta^4}\epsilon_{\mu\nu}.
\end{array}
\end{equation}
Note that they are $\alpha$-independent. Upon quantizing we get four observables obeying $\alpha$-independent algebra.
Any other observable including energy can be constructed out of them so its spectrum and other properties do not depend on $\alpha$.

\subsection{The case of third derivatives\label{subsec:2}}
Let us consider a nonsingular Lagrangian of the form
\begin{equation}
\label{l48}
L=L(q,\dot q,\overset{..}{q},\overset{...}{q}).
\end{equation}
It is slightly surprising that this case (and, in general, the case when the highest time derivatives are of odd
order - see Appendix) is simpler. We define the new variables 
\begin{equation}
q^{\mu}=q_1^{\mu},\quad \dot q^{\mu}=\dot q_1^{\mu},\quad \overset{..}{q}^{\mu}=q_2^{\mu},\quad {\overset{...}{q}}^{\mu}=\dot q_2^{\mu}.
\end{equation}
Next, the function $F(q_1,\dot q_1,q_2,q_3)$ is selected which obeys 
\begin{equation}
\label{l49}
\det\left(\frac{\partial^2 F}{\partial\dot q_1^{\mu}\partial q_3^{\nu}}\right)\neq 0;
\end{equation}
here $q_3^\mu$ are additional variables. The modified Lagrangian reads
\begin{eqnarray}
\label{l50}
\mL(q_1,\dot q_1,q_2,\dot q_2,q_3,\dot q_3)&=&L(q_1,\dot q_1,q_2,\dot q_2)+\frac{\partial F(q_1,\dot q_1,q_2,q_3)}{\partial q_1^{\mu}}\dot q_1^{\mu}+
\frac{\partial F(q_1,\dot q_1,q_2,q_3)}{\partial q_2^{\mu}}\dot q_2^{\mu}
\nonumber\\ &+& \frac{\partial F(q_1,\dot q_1,q_2,q_3)}{\partial q_3^{\mu}}\dot q_3^{\mu}+\frac{\partial F(q_1,\dot q_1,q_2,q_3)}{\partial\dot q_1^{\mu}}q_2^{\mu}.
\end{eqnarray}
It can be easily shown that the Euler-Lagrange equations for $\mL$ yield the initial equations for the original
variable $q^\mu\equiv q_1^\mu$. 
Again, as in the second-order case, the Legendre transformation can be directly performed due to the 
condition (\ref{l49}). The momenta read
\begin{eqnarray}
\label{l53}
&&p_{1\mu}=\frac{\partial L}{\partial \dot q_1^{\mu}}+\frac{\partial^2 F}{\partial q_1^{\nu}\partial \dot q_1^{\mu}}\dot q_1^{\nu}
+\frac{\partial^2 F}{\partial\dot q_1^{\mu}\partial \dot q_1^{\nu}} q_2^{\nu}+\frac{\partial^2 F}{\partial\dot q_1^{\mu}\partial  q_2^{\nu}}\dot q_2^{\nu}
+\frac{\partial^2 F}{\partial\dot q_1^{\mu}\partial  q_3^{\nu}}\dot q_3^{\nu}+\frac{\partial F}{\partial q_1^{\mu}},
\\
&& \label{l52}
p_{2\mu}=\frac{\partial L}{\partial \dot q_2^{\mu}}+\frac{\partial F}{\partial q_2^{\mu}},
\\
&& \label{l51}
p_{3\mu}=\frac{\partial F(q_1,\dot q_1,q_2,q_3)}{\partial q_3^{\mu}}.
\end{eqnarray}
By virtue of (\ref{l49}) one can solve (\ref{l51}) for $\dot q_1^\mu$,
\begin{equation}
\dot q_1^{\mu}=\dot q_1^{\mu}(q_1,q_2,q_3,p_3).
\end{equation}
Inserting  this solution into eq. (\ref{l52}) one computes
\begin{equation}
\dot q_2^{\mu}=\dot q_2^{\mu}(q_1,q_2,q_3,p_2,p_3);
\end{equation}
the solution is (at least locally) unique because $L$ is, by assumption, nonsingular in Ostrogradski sense.
Similarly, (\ref{l53}) can be solved in terms of $\dot q_3^\mu$:
\begin{equation}
\dot q_3^{\mu}=\dot q_3^{\mu}(q_1,q_2,q_3,p_1,p_2,p_3).
\end{equation}
Finally, Hamiltonian is of the form
\begin{equation}
\label{l54}
H=p_{1\mu}\dot q_1^{\mu}+p_{2\mu}\dot q_2^{\mu}-L-\frac{\partial F}{\partial q_1^{\mu}}\dot q_1^{\mu}-\frac{\partial F}{\partial \dot q_1^{\mu}}q_2^{\mu}
-\frac{\partial F}{\partial q_2^{\mu}}\dot q_2^{\mu},
\end{equation}
where everything is expressed in terms of $q$'s and $p$'s (the terms containing $\dot q_3^\mu$ cancel).
As above, we have checked that the canonical equations of motion yield the initial equation. 
The canonical transformation which relates our formalism to the Ostrogradski one reads
\begin{equation}
\label{l54b}
\begin{array}{l}
\tilde q_1^{\mu}=q_1^{\mu},\\
\tilde q_2^{\mu}=f^{\mu}(q_1,q_2,q_3,p_3),\\
\tilde q_3^{\mu}=q_2^{\mu},\\
\tilde p_{1\mu}=p_{1\mu}-\frac{\partial F}{\partial q_1^{\mu}}(q_1,f(q_1,q_2,q_3,p_3),q_2,q_3),\\
\tilde p_{2\mu}=-\frac{\partial F}{\partial f^{\mu}}(q_1,f(q_1,q_2,q_3,p_3),q_2,q_3),\\
\tilde p_{3\mu}=p_{2\mu}-\frac{\partial F}{\partial q_2^{\mu}}(q_1,f(q_1,q_2,q_3,p_3),q_2,q_3),\\
\end{array}
\end{equation}
where $f^{\mu}$  is the solution of eq. (\ref{l51}), i.e., $f^{\mu}=\dot q_1^{\mu}$.
The relevant generating function reads 
\begin{equation}
\Phi(q_1,\tilde p_1,q_2,\tilde q_2,q_3,\tilde p_3)=q_1^\mu\tilde p_{1\mu}+q_2^\mu\tilde p_{3\mu}
+F(q_1,\tilde q_2,q_2,q_3).
\end{equation}
Again, the advantage of our Hamiltonian  over the Ostrogradski one is that the former is nonsingular
in the sense that the inverse Legendre transformation can be performed directly. 
\subsection{The second order Lagrangian once more}
By comparing Section \ref{subsec:1} and \ref{subsec:2} we see that the modified Hamiltonian formalism
is  somewhat simpler in the case of third order Lagrangian (actually, as it is shown in Appendix,
this is the case for all Lagrangians of odd order). Namely, in latter case no counterpart of the condition 
(\ref{l40})  is necessary. This will appear to play the crucial role in the case of singular (in Ostrogradski sense)
Lagrangians (see Section \ref{sec:3} below). Therefore, as a preliminary step, we consider here the second
order Lagrangians as a special, singular case of third order ones. The resulting Hamiltonian formalism 
is then constrained. However, with an additional assumption that the function $F$ does not depend on
$q_2^\mu$, one can perform complete reduction of phase space obtaining  the structure described
in Section \ref{subsec:1}.
\par
Let 
\begin{equation}
L=L(q,\dot q,\overset{..}{q}),
\end{equation}
and $F=F(q_1,\dot q_1,q_3)$ obeys (\ref{l49}). We define 
\begin{eqnarray}
\mL(q_1,\dot q_1,q_2,q_3,\dot q_3)&=&L(q_1,\dot q_1,q_2)+\frac{\partial F(q_1,\dot q_1,q_3)}{\partial q_1^{\mu}}\dot q_1^{\mu}
\nonumber\\&+& \frac{\partial F(q_1,\dot q_1,q_3)}{\partial q_3^{\mu}}\dot q_3^{\mu}+\frac{\partial F(q_1,\dot q_1,q_3)}{\partial\dot q_1^{\mu}}q_2^{\mu}.
\end{eqnarray}
The relevant momenta read
\begin{eqnarray}
&&\label{l59}
p_{1\mu}=\frac{\partial L}{\partial \dot q_1^{\mu}}+\frac{\partial^2 F}{\partial q_1^{\nu}\partial \dot q_1^{\mu}}\dot q_1^{\nu}
+\frac{\partial^2 F}{\partial\dot q_1^{\mu}\partial \dot q_1^{\nu}} q_2^{\nu}
+\frac{\partial^2 F}{\partial\dot q_1^{\mu}\partial  q_3^{\nu}}\dot q_3^{\nu}+\frac{\partial F}{\partial q_1^{\mu}},
\\
&& \label{l58}
p_{2\mu}=0,
\\
&&
\label{l57}
p_{3\mu}=\frac{\partial F}{\partial q_3^{\mu}}.
\end{eqnarray}
There is one set of primary constraints (\ref{l58}). On the other hand, due to the condition (\ref{l49})
 $\dot q_1^\mu$ and $\dot q_3^\mu$ can be expressed in terms of $q_1,q_2,q_3,p_1,p_3$.
The Dirac Hamiltonian takes the form 
\begin{equation}
\label{l60}
H=p_{1\mu}\dot q_1^{\mu}-L-\frac{\partial F}{\partial q_1^{\mu}}\dot q_1^{\mu}
-\frac{\partial F}{\partial\dot q_1^{\mu}}q_2^{\mu}+c^\mu p_{2\mu},
\end{equation}
where $c^\mu$ are Lagrange multipliers enforcing the constraints $\Phi_{1\mu}\equiv p_{2\mu}\approx 0$.
\par
The stability of primary constraints implies 
\begin{equation}
\label{l67}
0\approx \dot\Phi_{1\mu}\equiv \Phi_{2\mu}=\frac{\partial L(q_1,\dot q_1(q_1,q_3,p_3),q_2)}{\partial q_2^{\mu}}+
\frac{\partial F(q_1,\dot q_1(q_1,q_3,p_3),q_3)}{\partial \dot q_1^{\mu}}.
\end{equation}
In order to check the stability of secondary constraints $\Phi_{2\mu}$  we note that, as it can be verified by direct computation, 
\begin{equation}
\label{l67b}
\{\dot q_1^\mu,\dot q_1^\nu\}=0.
\end{equation}
Using (\ref{l67b})  together with
\begin{equation}
0\approx \dot\Phi_{2\mu}=\{\Phi_{2\mu},H\},
\end{equation}
we arrive at the following condition
\begin{equation}
\label{l68}
\frac{\partial^2 L}{\partial q_2^{\mu}\partial q_2^{\nu}}c^{\nu}+\frac{\partial^2 L}{\partial q_1^{\mu}\partial q_1^{\nu}}\dot q_1^{\nu}+
\frac{\partial^2 L}{\partial q_2^{\mu}\partial\dot q_1^{\nu}}q_2^{\nu}+p_{1\mu}-\frac{\partial L}{\partial \dot q_1^{\mu}}-\frac{\partial F}{\partial q_1^{\mu}}=0.
\end{equation}
The initial Lagrangian is nonsingular and eq. (\ref{l68}) can be used to determine the Lagrange multipliers
$c^\nu$  uniquely. Therefore, the are no further constraints.
\par
In order to convert our constraints into strong equations we define Dirac brackets. To this end we compute
\begin{equation}
\{\phi_{1\mu},\phi_{1\nu}\} =0,
\end{equation}
\begin{equation}
\label{l68c}
\{\phi_{1\mu},\phi_{2\nu}\} =-\frac{\partial ^2 L}{\partial q_2^{\mu}\partial q_2^{\nu}}\equiv -W_{\mu\nu}.
\end{equation}
Moreover, 
\begin{equation}
\left\{ \frac{\partial L}{\partial q_{2}^{\mu}},\frac{\partial L}{\partial q_{2}^{\nu}}\right\}=0,\quad
 \left\{ \frac{\partial F}{\partial \dot q_{1}^{\mu}},\frac{\partial F}{\partial \dot q_{1}^{\nu}}\right\}=0,\quad 
\left\{ \frac{\partial F}{\partial \dot q_{1}^{\mu}},\frac{\partial L}{\partial q_{2}^{\nu}}\right\}=\frac{\partial^2 L}{\partial q_2^{\nu}{\partial\dot q_1^{\mu}}},
\end{equation}
which implies
\begin{equation}
\label{l72}
\{\phi_{2\mu},\phi_{2\nu}\}=\frac{\partial^2 L}{\partial \dot q_1^{\mu}\partial q_2^{\nu}}-\frac{\partial^2 L}{\partial \dot q_1^{\nu}\partial q_2^{\mu}}\equiv V_{\mu\nu}.
\end{equation}
By assumption, $W$ is a nonsingular matrix. Consequently, 
\begin{equation}
C=\left(
\begin{array}{cc}
\{\phi_{1\mu},\phi_{1\nu}\} &\{\phi_{1\mu},\phi_{2\nu}\} \\
\{\phi_{2\mu},\phi_{1\nu}\} &\{\phi_{2\mu},\phi_{2\nu}\} 
\end{array}
\right)=\left(
\begin{array}{cc}
0&-W\\
W&V
\end{array}\right),
\end{equation}
is also nonsingular and
\begin{equation}
C^{-1}=\left(
\begin{array}{cc}
W^{-1}VW^{-1}&W^{-1}\\
-W^{-1}&0
\end{array}\right).
\end{equation}
Dirac bracket takes the following form 
\begin{eqnarray}
\label{l64b}
\{\cdot ,\cdot\}_D&=&\{\cdot,\cdot\}-\{\cdot,\phi_{1\mu}\}(W^{-1}VW^{-1})_{\mu\nu}\{\phi_{1\nu},\cdot\}\nonumber\\
& &-\{\cdot,\phi_{1\mu}\}(W^{-1})_{\mu\nu}\{\phi_{2\nu},\cdot\}+\{\cdot,\phi_{2\mu}\}(W^{-1})_{\mu\nu}\{\phi_{1\nu},\cdot\}.
\end{eqnarray}
The  constraints $\Phi_{1\mu}$ depend on $p_{2\mu}$ only. We conclude from (\ref{l64b})
that the Dirac brackets  for $q_1^\mu,q_3^\mu,p_{1\mu},p_{3\mu}$ take the canonical form.
Moreover, $p_{2\mu}=0$  while $q_2^\mu$ can be determined from (\ref{l67}). Note that the
solution for $q_2^\mu$, by virtue of eq. (\ref{l40}) reads 
\begin{equation}
q_2^\mu=\chi^\mu(q_1,\dot q_1(q_1,q_3,p_3),q_3).
\end{equation}
So, up to renumbering $q_2\leftrightarrow q_3$ we arrived at the same scheme as in Section \ref{subsec:1}.
\par
In order to illustrate  the above approach, we use the same example as before:
\begin{equation}
L=\lambda \epsilon _{\mu\nu}\dot q^{\mu}{\overset{..}{q}}^{\nu}+\frac{\beta}{2}({\overset{..}{q}}^{\mu})^2,\quad \beta\neq 0,
\end{equation}
and
\begin{equation}
F=\alpha \dot q_1^{\mu} q_3^{\mu},\quad \alpha\neq 0.
\end{equation}
Then $H$ takes the form
\begin{equation}
\label{l65}
H=\frac{1}{\alpha}p_{1\mu}p_{3\mu}-\frac{\lambda}{\alpha}\epsilon_{\mu\nu}p_{3\mu}q_2^{\nu}-
\frac{\beta}{2}(q_2^{\mu})^2-\alpha q_2^{\mu}q_3^{\nu}+\frac{2\lambda}{\beta}
\epsilon_{\mu\nu}p_{2\mu}q_2^{\nu}-\frac{1}{\beta}p_{1\mu}p_{2\mu},
\end{equation}
while the constraints are
\begin{equation}
\phi_{1\mu}=p_{2\mu},\quad \phi_{2\mu}=-\frac{\lambda}{\alpha}\epsilon_{\mu\nu}p_{3\nu}+\beta q_2^{\mu}+\alpha q_3^{\mu},
\end{equation}
and serve to eliminate $p_{2\mu}$ and $q_2^\mu$,
\begin{equation}
p_{2\mu}=0,\quad q_2^{\mu}=\frac{\lambda}{\alpha\beta}\epsilon_{\mu\nu}p_{3\nu}-\frac{\alpha}{\beta} q_3^{\mu}.
\end{equation}
Inserting this back into the Hamiltonian we arrive at the following expression
\begin{equation}
H=\frac{1}{\alpha}p_{1\mu}p_{3\mu}+\frac{\lambda^2}{2\alpha^2\beta}(p_{3\mu})^2+
\frac{\lambda}{\beta}\epsilon_{\mu\nu}p_{3\mu}q_3^{\nu}+\frac{\alpha^2}{2\beta}(q_3^{\mu})^2
\end{equation}
which coincides with the one given by  eq. (\ref{l48b}) provided the replacement $q_2\leftrightarrow q_3$, $p_2\leftrightarrow p_3$ has been made.
\section{Singular Lagrangians of the second order \label{sec:3}}
In this section we consider the second order Lagrangians
\begin{equation}
L=L(q,\dot q,\overset{..}{q}),
\end{equation}
which are singular in the Ostrogradski sense, i.e.
\begin{equation}
\label{l68b}
\det(W_{\mu\nu})\equiv\det\left(\frac{\partial^2L}{\partial{\overset{..}{q}}^\mu\partial{\overset{..}{q}}^\nu}\right)=0.
\end{equation}
For standard  Ostrogradski approach to such singular Lagrangians see, for example, Ref. \cite{16b,17b}.
\par The formalism of Section \ref{subsec:1} is not directly applicable because due to eq. (\ref{l68b}),
eqs. (\ref{l40}) cannot be solved to determine the functions $\chi^\mu$. Moreover, eqs. (\ref{l40}) put
in this case further restrictions on the form of $F$.
\par
In order to get rid of these problems we will follow the method of  Section \ref{subsec:2} and consider $L$ as a third order singular Lagrangian. From this point of
view  its singularity comes both from eq. (\ref{l68b})  and the fact  that the third order time derivatives
 are absent. Given a singular Lagrangian $L$ we select a function $F=F(q_1,\dot q_1,q_3)$ obeying
(\ref{l49})  and define
\begin{eqnarray}
\mL(q_1,\dot q_1,q_2,q_3,\dot q_3)&=&L(q_1,\dot q_1,q_2)+\frac{\partial F(q_1,\dot q_1,q_3)}{\partial q_1^{\mu}}\dot q_1^{\mu}
\nonumber\\&+& \frac{\partial F(q_1,\dot q_1,q_3)}{\partial q_3^{\mu}}\dot q_3^{\mu}+
\frac{\partial F(q_1,\dot q_1,q_3)}{\partial\dot q_1^{\mu}}q_2^{\mu}.
\end{eqnarray}
As before, the canonical momenta given by (\ref{l58}) provide the primary constraints while 
(\ref{l59}) and (\ref{l57})  allow us  to compute $\dot q_1^\mu$ and $\dot q_3^\mu$.
The Hamiltonian is given by eq. (\ref{l60}). The secondary constraints  read again
\begin{equation}
\label{l70}
0\approx  \Phi_{2\mu}=\frac{\partial L(q_1,\dot q_1(q_1,q_3,p_3),q_2)}{\partial q_2^{\mu}}+
\frac{\partial F(q_1,\dot q_1(q_1,q_3,p_3),q_3)}{\partial \dot q_1^{\mu}}.
\end{equation}
Now we have to investigate the stability of $\Phi_{2\mu}$. To this end we assume  that $W$ has rank $K$, $K<N$;
this implies the existence of $J=N-K$ linearly independent null eigenvectors
$\gamma_a^\mu(q_1,\dot q_1,q_2)$, $a=1,2,\ldots,J$,
\begin{equation}
\label{l70b}
W_{\mu\nu}\gamma_a^\nu=0.
\end{equation}
Equations  (\ref{l68})  do not determine uniquely the Lagrange multipliers $c^\mu$; on the contrary,
we get new  constraints of the form
\begin{equation}
\label{l71}
0\equiv \Phi_{3a}=\gamma_a^{\mu}\left(\frac{\partial^2 L}{\partial q_1^{\mu}\partial q_1^{\nu}}\dot q_1^{\nu}+
\frac{\partial^2 L}{\partial q_2^{\mu}\partial\dot q_1^{\nu}}q_2^{\nu}+
p_{1\mu}-\frac{\partial L}{\partial \dot q_1^{\mu}}-\frac{\partial F}{\partial q_1^{\mu}}\right);
\end{equation}
here, as previously,  $\dot q_1^\mu=\dot q_1^\mu(q_1,q_3,p_3)$, so the above constraints contain 
$q_1,q_2,q_3,p_1$ and $p_3$.
\par 
We have started with third order formalism; therefore, our phase space is $6N$-dimensional.
As in  nonsingular case (Section \ref{sec:2}) we would like to eliminate $q_2$'s and $p_2$'s.
The latter are  equal zero by primary constraints $\Phi_{1\mu}$.
As far  as $q_2$'s are considered the situation  is more involved.
\par
First, by virtue  of the assumption (\ref{l68b}) about $W$ we can determine from eqs.  (\ref{l70})
K variables $q_2^\mu$ in terms of $q_1,p_1,q_3,p_3$ and the remaining $q_2$'s. By substituting the resulting
expression back to eqs. (\ref{l70}) we arrive at $J$ constraints on  $q_1,p_1,q_3$ and $p_3$. We denote these new
constraints  by $\psi_a(q_1,q_3,p_1,p_3)$.
Let us now concentrate on the constraints (\ref{l71}). In general, they contain the $q_2^\mu$
variables and imply the constraints  on $q_1,q_3,p_1,p_3$ only provided $q_2$'s enter in the combinations
which can be determined from eqs. (\ref{l70}). In order to decide if it happens consider the variations 
$\delta q_2^\mu$ which do not change the RHS of (\ref{l70}). From the definition of $W_{\mu\nu}$
 we conclude that such $\delta q_2^\mu$  are linear combinations of $\gamma_a^\mu$ (see  (\ref{l70b})).
If the RHS of (\ref{l71}) are stationary under such variations $\delta q_2^\mu$, eqs. (\ref{l70}) and
(\ref{l71}) can be combined  to yield the constraints  which do not depend on $q_2$'s. The relevant condition
reads
\begin{equation}
\frac{\partial \Phi_{3a}}{\partial q_2^{\mu}}\gamma_{b}^{\mu}=0,\quad b=1,\ldots,J;
\end{equation}
where  $a$ takes $M$ values, which without loss of generality can be chosen as $a=1,\ldots, M$.
In this way we obtain $M$ new constraints on $q_1,p_1,q_3,p_3$.
\par
One can check that  
\begin{equation}
\frac{\partial \Phi_{3a}}{\partial q_2^{\mu}} \gamma^{\mu}_b=\gamma_b^{\mu}\gamma_{a}^{\nu}\left(\frac{\partial^2 L}{\partial \dot q_1^{\mu}\partial q_2^{\nu}}-
\frac{\partial^2 L}{\partial \dot q_1^{\nu}\partial q_2^{\mu}} \right).
\end{equation}
By virtue of   (\ref{l72})  we find
\begin{equation}
\{\psi_{a},\psi_{b}\}\approx \gamma_a^{\mu}\gamma_b^{\nu}\{\phi_{2\mu},\phi_{2\nu}\}, \quad a=1,\ldots,M,\, b=1,\ldots,J.
\end{equation}
Let us summarize. For the nonsingular second order Lagrangian  viewed as a singular third order one,
$(q_1,p_1,q_3,p_3)$ forms the reduced phase space; no further constraints exist. On the contrary, in the 
singular case $q_1,p_1,q_3,p_3$ are still constrained. First, there exist $J$ constraints
$\psi_a(q_1,p_1,q_3,p_3)$; moreover, if some (say - $M$) $\psi$'s are in involution (on the constraint 
surface) with all $\psi$'s  there exist additional $M$ constraints following from  eqs. (\ref{l70}) and (\ref{l71}).
This agrees  with the  conclusions of Ref. \cite{16b}.
\par
In general, for singular Lagrangian it is not possible to determine uniquely all Lagrange multipliers $c^\mu$.
However, we are in fact interested  only in dynamical equations for  $q_{1},q_3,p_{1}$ and $p_3$.
Therefore, we can use the following Hamiltonian  
\begin{equation}
\label{l75}
H=p_{1\mu}\dot q_1^{\mu}-L-\frac{\partial F}{\partial q_1^{\mu}}\dot q_1^{\mu}
-\frac{\partial F}{\partial\dot q_1^{\mu}}q_2^{\mu}.
\end{equation}
On the constraint surface it does not depend on $q_2$'s, 
\begin{equation}
\frac{\partial H}{\partial q_2^{\mu}}=-\frac{\partial L}{\partial q_2^{\mu}}-\frac{\partial F}{\partial \dot q_1^{\mu}}\approx 0.
\end{equation}
The existence of further secondary constraints depend on the particular form of the Lagrangian.
\par
Finally, let us note that the canonical transformation (\ref{l54b}) leads to the form of dynamics presented 
in Ref. \cite{16b}. However, within our procedure the Legendre transformation  from the tangent bundle
 of configuration manifold to phase manifold  is again straightforward (if one takes into account standard
modifications due to the existence of constraints).

Singular higher derivative Lagrangians were also considered in~\cite{18r}. The authors considered the physically important case of 
reparametrization invariant theories (higher-derivative reparametrization invariant Lagrangians appear, for example, in the description
of radiation reaction~\cite{19r}). In their geometrical approach the image of the Legendre transformation form a submanifold of some cotangent bundle.
This suggests that in the case of higher-derivative singular theories it is advantageous to start with enlarged phase space; this agrees 
with our conclusions. 

To conclude this section with a simple example consider the following  Lagrangian 
\begin{equation}
\label{l75b}
L=\lambda\epsilon_{\mu\nu}\dot q^{\mu}{\overset{..}{q}}^{\nu}+\frac{\beta}{2}({\overset{..}{q}^1})^2,  \quad \mu,\nu=1,2.
\end{equation}
It is singular and the matrix $W$ (eq.  (\ref{l68c})) is of rank $1$ for $\beta\neq 0$ and $0$ for $\beta=0$.
We take  $F$ as
\begin{equation}
F=\alpha\dot q_1^{\mu}q_3^{\mu}.
\end{equation}
Assume first $\beta\neq 0$. Then
\begin{equation}
\mL=\lambda \epsilon_{\mu\nu}\dot q_1^{\mu}q_2^{\nu}+\frac{\beta}{2}(q_2^1)^2+\alpha q_3^{\mu}q_2^{\mu}+\alpha\dot q_1^{\mu}\dot  q_3^{\mu},
\end{equation}
and
\begin{equation}
\begin{array}{l}
p_{1\mu}=\lambda\epsilon_{\mu\nu}q_2^{\nu}+\alpha\dot q_3^{\mu},\\
p_{2\mu}=0,\\
p_{3\mu}=\alpha\dot q_1^{\mu}.
\end{array}
\end{equation}
The primary constraints are
\begin{equation}
\Phi_{1\mu}=p_{2\mu}\approx 0;
\end{equation}
while the Hamiltonian reads
\begin{equation}
H=\frac{1}{\alpha}p_{1\mu}p_{3\mu}-\frac{\lambda}{\alpha}\epsilon_{\mu\nu}p_{3\mu}q_{2}^{\nu}-\frac{\beta}{2}(q_2^1)^2-\alpha q_2^{\mu}q_3^{\mu}
+c^\mu p_{2\mu}.
\end{equation}
One easily derives the secondary constraints 
\begin{equation}
\begin{array}{l}
0\approx\Phi_{21}=\frac{\lambda}{\alpha}p_{32}-\beta q_2^1-\alpha q_3^1,\\
0\approx\Phi_{22}=\frac{\lambda}{\alpha}p_{31}+\alpha q_3^2.
\end{array}
\end{equation}
The stability for $\Phi_{2\mu}$  yields 
\begin{equation}
\label{l75c}
0 \approx \{\Phi_{21},H\}={2\lambda}q_2^2-\beta c^1-p_{11},
\end{equation}
\begin{equation}
\label{l75d}
0 \approx \{\Phi_{22},H\}=p_{12}+2\lambda q_2^1=\Phi_3.
\end{equation}
Equation (\ref{l75c}) allows us to compute $c^1$,
\begin{equation}
\label{l75e}
c^1=\frac{1}{\beta}(2\lambda q_2^2-p_{11}),
\end{equation}
while (\ref{l75d}) provides a new constraint. Its stability enforces $c^1=0$ which together with (\ref{l75e})
 yields further constraint
\begin{equation}
0\approx \Phi_4=\frac{1}{\beta}(2\lambda q_2^2-p_{11}).
\end{equation}
Finally, differentiating the above equation with respect to time we get $c^2=0$.
The resulting  Hamiltonian is  
\begin{equation}
H=\frac{1}{\alpha}p_{1\mu}p_{3\mu}-\frac{\lambda}{\alpha}\epsilon_{\mu\nu}p_{3\mu}q_{2}^{\nu}-
\frac{\beta}{2}(q_2^1)^2-\alpha q_2^{\mu}q_3^{\mu}.
\end{equation}
Still we have  to take into account the constraints $\Phi_{2\mu},\Phi_3$ and $\Phi_4$. The latter two
can be rewritten as
\begin{equation}
\Phi_{3\mu}=p_{1\mu}-2\lambda\epsilon_{\mu\nu}q_2^{\nu}.
\end{equation}
$\Phi_{2\mu}$  and $\Phi_{3\mu}$ are now used in order to eliminate all variables except  $q_1^\mu,p_{1\mu}$
and $q_3^2,p_{32}$.
The only nonstandard Dirac bracket reads 
\begin{equation}
\{q_3^2,p_{32}\}_D=\frac{1}{2}.
\end{equation}
The Hamiltonian, when expressed in terms of unconstrained variables, takes the form
\begin{equation}
H=\frac{1}{\alpha}p_{12}p_{32}-\frac{\alpha}{\lambda}p_{11}q_3^2+\frac{\beta}{8\lambda^2}(p_{12})^2.
\end{equation}
Let us note that the limit $\beta\rightarrow 0$ is smooth. Of course, we could put $\beta=0$ from the very
beginning  and arrive at the same conclusion.
\section{An example: mini-superspace formulation of $f(R)$ gravity \label{sec:4}}
As a more elaborate but still a toy example we consider mini-superspace Hamiltonian formulation of $f(R)$ gravity \cite{11b}.
We consider the following  (LFRW - type) metrics 
\begin{equation}
ds^2=-N^2dt^2+a^2d{\vec x}^2.
\end{equation}
Under such reduction the Lagrangian of $f(R)$ gravity takes the form
\begin{equation}
L(a,N)=\frac{1}{2}Na^3f(R),
\end{equation}
where the curvature is given by
\begin{equation}
R=6{\left(\frac{\dot a}{NA}\right)}^.+12{\left(\frac{\dot a}{Na}\right)}^2.
\end{equation}
We see that $L$ depends on second time derivatives. We proceed along the lines described in Section \ref{sec:2}.
The basic dynamical variables are chosen as follows 
\begin{equation}
a_1=a,\;\dot a_1=\dot a,\; N_1=N,\; \dot N_1=\dot N,\; a_2=R,
\end{equation}
while 
\begin{equation}
\overset{..}{a}=\chi(a_1,\dot a_1,N_1,\dot N_1,a_2),
\end{equation}
is determined by eq. (\ref{l40}) once appropriate $F$ is selected. We take
\begin{equation}
\label{l76}
F=-3a^2_1f'(a_2)\dot a_1;
\end{equation}
under the assumption $f''\neq 0$, eqs. (\ref{l76}) and (\ref{l40}) yield
\begin{equation}
\label{l77}
a_2=R.
\end{equation}
Solving (\ref{l77})  with respect to $\overset{..}{a}$ we find 
\begin{equation}
\overset{..}{a}=\frac{a_1N_1}{6}\left(R-\frac{6}{N_1^2a_1^2}((2-N_1){\dot a}_1^2-a_1{\dot a}_1{\dot N}_1)\right).
\end{equation}
The modified  Lagrangian reads
\begin{eqnarray}
\mL&=&\frac{1}{2}a_1^3N_1f(a_2)+f'(a_2)\left(-9a_1{\dot a}_1^2+\frac{6 a_1{\dot a_1}^2}{N_1}-
\frac{1}{2}a_1^3N_1a_2-\frac{3a_1^2{\dot a}_1{\dot N}_1}{N_1}\right)\nonumber\\
& -&3f''(a_2)a_1^2{\dot a}_1{\dot a}_2.
\end{eqnarray}
It is  straightforward to check that $\mL$ leads to the correct equations of motion. In order to simplify our
considerations  we introduce new variable
\begin{equation}
n_1=N_1f'(a_2).
\end{equation}
In terms of new variable $\mL$  reads
\begin{eqnarray}
\mL&=&\frac{1}{2}a_1^3n_1\frac{f(a_2)}{f'(a_2)}-9a_1{\dot a}_1^2f'(a_2)+\frac{6 a_1{\dot a_1}^2{f'}^2(a_2)}{n_1}
\nonumber\\
& -& \frac{1}{2}a_1^3n_1a_2-\frac{3a_1^2{\dot a}_1{\dot n}_1f'(a_2)}{n_1}.
\end{eqnarray}
Now, we compute the canonical momenta:
\begin{eqnarray}
&&p_1\equiv \frac{\partial \mL}{\partial{\dot n}_1}=-\frac{3a_1^2}{n_1}f'(a_2)\dot a_1 \label{l78}, \\
&&\pi_1\equiv \frac{\partial \mL}{\partial{\dot a}_1}=-18a_1{\dot a}_1f'(a_2)+\frac{12a_1{\dot a}_1{f'}^2(a_2)}{n_1}
-3\frac{a_1^2}{n_1}f'(a_2){\dot n}_1 \label{l79},\\
&&\pi_2\equiv \frac{\partial \mL}{\partial{\dot a}_2}=0.
\end{eqnarray}
One can solve (\ref{l78}) and (\ref{l79}) in terms of $\dot a_1$ and $\dot  n_1$. We form the Hamiltonian
\begin{eqnarray}
H&=&-\frac{n_1p_1\pi_1}{3a_1^2f'(a_2)}-\frac{n_1a_1^3f(a_2)}{2f'(a_2)}+\frac{n_1^2p_1^2}{a_1^3f'(a_2)}-
\frac{2n_1p_1^2}{3a_1^3}\nonumber\\
&+&\frac{1}{2}a_1^3a_2n_1+\mu\pi_2\equiv \tilde H+\mu \pi_2.
\end{eqnarray}
Now, we investigate the stability of $\Phi_1\equiv \pi_2$ constraint
\begin{equation}
0\approx \dot \Phi_1=\{\Phi_1,H\}=\frac{f''(a_2)}{f'(a_2)}\left(\tilde H+\frac{2n_1p_1^2}{3a_1^3}-
\frac{a_1^3a_2n_1}{2}\right)\equiv\Phi_2.
\end{equation}
The stability condition for $\Phi_2$ determines $\mu$; an explicit expression for $\mu$ 
is irrelevant for what follows. In fact, $(\Phi_1,\Phi_2)$ are second class  constraints
\begin{equation}
\{\Phi_1,\Phi_2\}\approx \frac{f''(a_2)a_1^3N_1}{2f'(a_2)}.
\end{equation}
Thus, the constraints can be solved provided we use Dirac brackets. In particular, the Hamiltonian takes a 
simple form
\begin{equation}
\label{l80}
H=\tilde H=\frac{1}{2}a_1^3a_2n_1-\frac{2}{3}\frac{n_1p_1^2}{a_1^3},
\end{equation}
where 
\begin{equation}
\label{l81}
a_2=f^{-1}\left(-\frac{2p_1\pi_1}{3a_1^5}+\frac{2n_1p_1^2}{a_1^6}\right).
\end{equation}
Moreover, Dirac brackets for the variables $a_1,n_1,\pi_1,p_1$ remain canonical. Therefore, eqs. (\ref{l80})
and (\ref{l81}) give the complete Hamiltonian description. We have checked explicitly that  it leads to correct 
equations of motion. In the case under consideration our formalism, when compared with Ostrogradski  version,
seems to be more complicated. However, it has an advantage that the curvature $R$ is one of basic variables.
\section{Field theory \label{sec:5}}
Our formalism has a straightforward generalization to the field theory case.
For simplicity, we consider only the Lagrangian densities  depending on first and second derivatives. Such a density can be written in the form
\begin{equation}
\mL=\mL(\Phi,\partial_k\Phi,\partial_k\partial_l\Phi,\dot \Phi,\partial_k\dot\Phi,\overset{..}{\Phi}).
\end{equation}
Again, we put $\Phi=\Phi_1$ and select a function $F=F(\Phi_1,\dot \Phi_1,\Phi_2)$ obeying
\begin{equation}
\label{fe:1}
\frac{\partial^2F}{\partial\dot\Phi_1\partial\Phi_2}\neq 0;
\end{equation}
in the case of multicomponent field the relevant matrix should be nonsingular.
We define, as previously, the function
\begin{equation}
\chi=\chi(\Phi_1,\partial_k\Phi_1,\partial_k\partial_l\Phi_1,\dot\Phi_1,\partial_k\dot\Phi_1,\Phi_2),
\end{equation}
as the (locally unique by virtue of (\ref{fe:1})) solution to the equation
\begin{equation}
\frac{\partial\mL(\Phi_1,\partial_k\Phi_1,\partial_k\partial_l\Phi_1,\dot\Phi_1,\partial_k\dot\Phi_1,\chi)}
{\partial\chi}+\frac{\partial F(\Phi_1,\dot \Phi_1,\Phi_2)}{\partial\dot \Phi_1}=0.
\end{equation}
Finally, the new Lagrangian density reads
\begin{align}
\tilde\mL&=\mL(\Phi_1,\partial_k\Phi_1,\partial_k\partial_l\Phi_1,\dot\Phi_1,\partial_k\dot\Phi_1,\chi(\ldots))+
\frac{\partial F(\Phi_1,\dot\Phi_1,\Phi_2)}{\partial \Phi_1}\dot\Phi_1\nonumber\\
&+\frac{\partial F(\Phi_1,\dot\Phi_1,\Phi_2)}{\partial \Phi_2}\dot\Phi_2+
\frac{\partial F(\Phi_1,\dot\Phi_1,\Phi_2)}{\partial\dot \Phi_1}\chi(\ldots).
\end{align}
It is now straightforward to check that the Lagrange equations
\begin{equation}
\frac{\partial \tilde L}{\partial \Phi_i}-\partial_k\frac{\partial\tilde \mL}{\partial(\partial_k\Phi_i)}+
\partial_k\partial_l\frac{\partial\tilde\mL}{\partial(\partial_k\partial_l\Phi_i)}-\frac{d}{dt}\left(\frac{\partial\tilde\mL}{\partial\dot\Phi_i}
-\partial_k\frac{\partial\tilde\mL}{\partial(\partial_k\dot\Phi_i)}\right)=0,
\end{equation}
yield the initial equation for the original variable $\Phi\equiv\Phi_1$; as in the Section \ref{subsec:1} $\overset{..}{\Phi}=\chi(\ldots)$.
One can now perform the Legendre  transformation. The canonical momenta read
\begin{equation}
\label{fe:2}
\pi_i(x)=\frac{\delta \tilde L}{\delta\dot \Phi_i(x)},\quad \tilde L\equiv\int d^3 x\tilde\mL.
\end{equation}
Equations  (\ref{fe:2}) can be solve (due to (\ref{fe:1})) with respect to $\dot\Phi_i$:
\begin{align}
&\dot\Phi_1=\dot\Phi_1(\Phi_1,\Phi_2,\Pi_2)\\
&\dot\Phi_2=\dot\Phi_2(\Phi_1,\partial_k\Phi_1,\partial_k\partial_l\Phi_1,\partial_k\partial_l\partial_m\Phi_1,\Phi_2,\partial_k\Phi_2,\partial_k\partial_l\Phi_2,
\Pi_1,\Pi_2,\partial_k\Pi_2,\partial_k\partial_l\Pi_2).
\end{align}
$H$ is defined in a standard way 
\begin{equation}
H=\int d^3x(\Pi_1(x)\dot\Phi_1(x)+\Pi_2(x)\dot\Phi_2(x))-\tilde L,
\end{equation}
and leads to the correct canonical equations of motions.
\appendix
\section*{Appendix}
\section{Extension to the case of arbitrary high derivatives \label{app:1}} 
Here we generalize the approach proposed in Section \ref{sec:2} to the case of Lagrangians containing  time derivatives
of arbitrary order \cite{18b}. We restrict ourselves to the  case of one degree of freedom.
We start with the Lagrangian depending on time derivatives up to some even order
\begin{equation}
L=L(q,\dot q,\ddot q,\ldots,q^{(2n)}),    \label{w1}
\end{equation}
which is assumed to be nonsingular in Ostrogradski sense, $\frac{\partial ^2L}{\partial {q^{{(2n)}^2}}}\not=0$.
Define new variables
\begin{eqnarray}
&& q_i\equiv q^{(2i-2)}, \quad i=1,\ldots,n+1,   \label{w2} \\
&& \dot q_i\equiv q^{(2i-1)}, \quad  i=1,\ldots,n,   \nonumber
\end{eqnarray}
so that 
\begin{equation}
L=L(q_1,\dot q_1,q_2,\dot q_2,\ldots,q_n,\dot q_n,q_{n+1}).  \label{w3}
\end{equation}
Let further $F$\ be any  function of the following variables
\begin{equation}
F=F(q_1,\dot q_1,\ldots,q_n,\dot q_n,q_{n+1},q_{n+2},\ldots,q_{2n}),    \label{w4}
\end{equation}
obeying
\begin{equation}
 \quad  \frac{\partial L}{\partial q_{n+1}}+\frac{\partial F}{\partial \dot q_n}=0, \label{w5} 
\end{equation}
and
\begin{equation}
  \quad  \det\left[\frac{\partial ^2F}{\partial q_i\partial \dot q_j}\right]_{\genfrac{}{}{0pt}{}{n+2\leq i\leq 2n} { 1\leq j\leq n-1}}\neq 0,\quad n\geq 2,   \label{w6}
\end{equation}
(for $n=1$\ only (\ref{w5}) remains).

Finally, we define a new Lagrangian
\begin{equation}
\mathcal{L}\equiv L+\sum\limits_{k=1}^n\left(\frac{\partial F}{\partial q_k}\dot q_k+\frac{\partial F}{\partial \dot q_k}q_{k+1}\right)
+\sum\limits_{j=n+1}^{2n}\frac{\partial F}{\partial q_j}\dot q_j.    \label{w7}
\end{equation}
Let us have a look on Lagrange equations
\begin{equation}
 \frac{\partial \mathcal{L}}{\partial q_i}-\frac{d}{dt}\left(\frac{\partial \mathcal{L}}{\partial \dot q_i}\right)=0, \quad  i=1,\ldots,2n.  \label{w8}
\end{equation}
Using (\ref {w3}), (\ref {w4}), (\ref {w7}) and (\ref {w8}) one finds 
\begin{equation}
\sum\limits_{k=1}^n\frac{\partial ^2F}{\partial q_i\partial \dot q_k}(q_{k+1}-\ddot q_k)=0 \quad  i=n+1,\ldots,2n.   \label{w9}
\end{equation}
Consider the matrix $\left[\frac{\partial ^2F}{\partial q_i\partial \dot q_j}\right]_{\genfrac{}{}{0pt}{}{n+1\leq i\leq 2n}{1\leq j\leq n}}$\
 entering the LHS of eq.(\ref{w9}). By virtue of (\ref{w5}), $\frac {\partial ^2F}{\partial q_i\partial \dot q_n}=0$\
 for $ i=n+2,\ldots,2n$\ while $\frac{\partial ^2F}{\partial q_{n+1}\partial \dot q_n}=- \frac{\partial ^2L}{\partial q_{n+1}^2}\not=0$\
 due to Ostrogradski nonsingularity condition. Therefore, the first column of our matrix has only one non vanishing element.
 This, together with the condition (\ref{w6}) implies that it is invertible. Therefore, eq. (\ref{w9}) gives
\begin{equation}
q_{k+1}=\ddot q_k, \quad  k=1,\ldots,n.   \label{w10}
\end{equation}
Let us now consider (\ref {w8}) for $1\leq i\leq n$. We find
\begin{equation}
\frac{\partial L}{\partial q_i}-\frac{d}{dt}\left(\frac{\partial L}{\partial \dot q_i}\right)+\frac{\partial F}{\partial \dot q_{i-1}}
-\frac{d^2}{dt^2}\left(\frac{\partial F}{\partial \dot q_i}\right)=0, \quad  i=1,\ldots,n,    \label{w11}
\end{equation}
where, by definition, $\frac{\partial F}{\partial \dot q_0}=0$. By combining these equations and using (\ref {w5}) and (\ref {w10}) we arrive 
finally at the initial Lagrange equation.
We conclude that, contrary to the case of Ostrogradski Lagrangian, our modified Lagrangian leads to proper equation of motion.
Let us now consider the Hamiltonian formalism. Again, the Legendre transformation can be immediately performed; neither 
additional Lagrange multipliers nor constraints analysis are necessary. In fact, let us define the canonical momenta in a standard way
\begin{equation}
p_i=\frac{\partial \mathcal{L}}{\partial \dot q_i},    \label {w13}
\end{equation}
so that
\begin{eqnarray}
p_i&=&\frac{\partial F}{\partial q_i}, \quad  i=n+1,\ldots,2n   \label {w14}\\
 p_i&=&\frac{\partial L}{\partial \dot q_i}+\sum\limits_{k=1}^n\left(\frac{\partial ^2F}{\partial \dot q_i\partial q_k}\dot q_k+\frac{\partial ^2F}
{\partial \dot q_i\partial \dot q_k}q_{k+1}\right) \nonumber \\
&+&\sum\limits_{j=n+1}^{2n}\frac{\partial ^2F}{\partial \dot q_i\partial q_j}\dot q_j+
\frac{\partial F}{\partial q_i}, \quad  i=1,\ldots,n.   \label {w15}
\end{eqnarray}
Due to nonsingularity of $\left[\frac{\partial ^2F}{\partial q_i\partial \dot q_j}\right]_{\genfrac{}{}{0pt}{}{n+1\leq i\leq 2n}{1\leq j\leq n}}$\
  eqs.(\ref {w14}) can be solved for $\dot q_1,\dot q_2,\ldots,\dot q_n$\
\begin{equation}
\dot q_i=f_i(q_1,\ldots,q_{2n},p_{n+1},\ldots,p_{2n}), \quad  i=1,\ldots,n.   \label {w16}
\end{equation}
Now, eqs.(\ref{w15}) are linear with respect to $\dot q_i, i=n+1,\ldots,2n$\ and can be easily solved. Finally, the Hamiltonian is 
calculated according to the standard prescription.

In order to compare the present formalism with the Ostrogradski approach let us note that they must be related by a canonical transformation. 
To see this we define new (Ostrogradski) variables $\tilde q_k,\tilde p_k, \; 1\leq k\leq 2n$:
\begin{align}
& \tilde q_{2i-1}=q_i, \quad  i=1,..,n,       \label {w17} \\
& \tilde q_{2i}=f_i(q_1,\ldots,q_{2n},p_{n+1},\ldots,p_{2n}), \quad  i=1,\ldots,n,    \label {w18} \\
& \tilde p_{2i-1}=p_i-\frac{\partial F}{\partial q_i}(q_1,f_1(\ldots),\ldots,q_n,f_n(\ldots),q_{n+1},\ldots,q_{2n}), \quad  i=1,\ldots,n,   \label {w19} \\
& \tilde p_{2i}=-\frac{\partial F}{\partial f_i}(q_1,f_1(\ldots),\ldots,q_n,f_n(\ldots),q_{n+1},\ldots,q_{2n}), \quad  i=1,\ldots,n.    \label {w20}
\end{align}
It is easily seen that the above transformation is a canonical one, i.e. the Poisson brackets are invariant. It is not hard to find the relevant 
generating function
\begin{eqnarray}
&& \Phi (q_1,\ldots,q_{2n},\tilde p_1,\tilde q_2,\tilde p_3,\tilde q_4,\ldots,\tilde p_{2n-1},\tilde q_{2n})    \label {w21} \\
&& =\sum\limits_{k=1}^nq_k\tilde p_{2k-1}+F(q_1,\tilde q_2,q_2,\tilde q_4,\ldots,q_n,\tilde q_{2n},q_{n+1},\ldots,q_{2n}).  \nonumber
\end{eqnarray}
\par
Let us now consider the case of Lagrangian depending on time derivatives up to some odd order
\begin{equation}
L=L(q,\dot q,\ddot q,\ldots,q^{(2n+1)}).    \label {w22}
\end{equation}
Again, we define
\begin{eqnarray}
&& q_i\equiv q^{(2i-2)}, \quad  i=1,\ldots,n+1,   \label {w23}  \\
&& \dot q_i\equiv q^{(2i-1)},  \quad  i=1,\ldots,n+1,   \label {w24}
\end{eqnarray}
so that
\begin{equation}
L=L(q_1,\dot q_1,q_2,\dot q_2,\ldots,q_{n+1},\dot q_{n+1}).     \label {w25}
\end{equation}
Now, we select a function $F$,
\begin{equation}
F=F(q_1,\dot q_1,q_2,\dot q_2,\ldots,q_n,\dot q_n,q_{n+1},\ldots,q_{2n+1}),    \label {w26}
\end{equation}
subject to the single condition
\begin{equation}
\det\left[\frac{\partial ^2F}{\partial q_i\partial \dot q_k}\right]_{\genfrac{}{}{0pt}{}{n+2\leq i\leq 2n+1}{1\leq k\leq n}}\neq 0,   \label {w27}
\end{equation}
and define the Lagrangian
\begin{equation}
\mathcal{L}=L+\sum\limits_{k=1}^n\left(\frac{\partial F}{\partial q_k}\dot q_k+\frac{\partial F}{\partial \dot q_k}q_{k+1}\right)+
\sum\limits_{j=n+1}^{2n+1}\frac{\partial F}{\partial q_j}\dot q_j  \label {w28}.
\end{equation}
Consider the Lagrange equations (\ref {w8}). First, we have
\begin{equation}
\sum\limits_{k=1}^n\frac{\partial ^2F}{\partial q_i\partial \dot q_k}(q_{k+1}-\ddot q_k)=0, \quad  i=n+2,\ldots,2n+1,   \label {w29}
\end{equation}
and, by virtue of (\ref {w27})
\begin{equation}
q_{k+1}=\ddot q_k, \quad  k=1,\ldots,n.   \label {w30}
\end{equation}
The remaining equations read
\begin{equation}
\frac{\partial L}{\partial q_i}-\frac{d}{dt}\left(\frac{\partial L}{\partial \dot q_i}\right)+\frac{\partial F}{\partial \dot q_{i-1}}-
\frac{d^2}{dt^2}\left(\frac{\partial F}{\partial \dot q_i}\right)=0, \quad  i=1,\ldots,n+1,    \label {w31}
\end{equation}
with $\frac{\partial F}{\partial \dot q_0}=0, \; \frac{\partial F}{\partial \dot q_{n+1}}=0$. 
Combining (\ref {w30}) and (\ref {w31}) one gets
\begin{equation}
\sum\limits_{k=0 }^{2n+1}(-1)^k\frac{d^k}{dt^k}\left(\frac{\partial L}{\partial q^{(k)}}\right)=0.    \label {w32}
\end{equation}
Let us note that no condition of the form (\ref {w5}) is here necessary.
\par
Also in the odd case the present formalism is related to that of Ostrogradski by a canonical transformation.
 Indeed, the canonical momenta read 
\begin{eqnarray}
p_i&=&\frac{\partial F}{\partial q_i}, \quad  i=n+2,\ldots,2n+1     \label {w33},  \\
p_i&=&\frac{\partial L}{\partial \dot q_i}+\sum\limits_{k=1}^n\left(\frac{\partial ^2F}{\partial \dot q_i\partial q_k}\dot q_k+
\frac{\partial ^2F}{\partial \dot q_i\partial \dot q_k}q_{k+1}\right)  \nonumber \\
&+&\sum\limits_{j=n+1}^{2n+1}\frac{\partial ^2F}{\partial \dot q_i\partial q_j}
\dot q_j,  \quad  i=1,\ldots,n+1;   \label{w55}
\end{eqnarray}
by virtue of (\ref{w27}) one can solve eqs. (\ref{w33}) for $\dot q_1,\ldots,\dot q_n$. The remaining $n+1$\ equations 
(\ref{w55}) are used to compute the velocities $\dot q_{n+1},\ldots,\dot q_{2n+1}$. 
In fact, using eqs. (\ref{w25}) -- (\ref{w27}) as well as the Ostrogradski nonsingularity condition 
$\frac{\partial ^2L}{\partial \dot q_{n+1}^2}\not=0$\ one easily finds
\begin{equation}
\det\left[\frac{\partial p_i}{\partial \dot q_i}\right]_{\genfrac{}{}{0pt}{}{1\leq i\leq n+1}{n+1\leq j\leq 2n+1}}\neq 0.\label{w56} 
\end{equation}
In particular
\begin{equation}
\dot q_i=f_i(q_1,\ldots,q_{2n+1},p_{n+2},\ldots,p_{2n+1}), \quad  i=1,\ldots,n.    \label {w34}
\end{equation}
Now, one can define the canonical transformation to Ostrogradski variables
\begin{align}
&\tilde q_{2i-1}=q_i, \quad  i=1,\ldots,n+1,   \label {w35}   \\
&\tilde q_{2i}=f_i(q_1,\ldots,q_{2n+1},p_{n+2},\ldots,p_{2n+1}), \quad  i=1,\ldots,n,    \label {w36} \\
&\tilde p_{2i-1}=p_i-\frac{\partial F}{\partial q_i}(q_1,f_1(\ldots),\ldots,q_n,f_n(\ldots),q_{n+1},\ldots,q_{2n+1}),  \;  i=1,\ldots,n+1,    \label {w37} \\
 &\tilde p_{2i}=-\frac{\partial F}{\partial f_i}(q_1,f_1(\ldots),\ldots,q_n,f_n(\ldots),q_{n+1},\ldots,q_{2n+1}), \quad  i=1,\ldots,n.   \label {w38}
\end{align}

The relevant generating function reads
\begin{eqnarray}
&& \Phi (q_1,q_2,\ldots,q_{2n+1},\tilde p_1,\tilde q_2,\tilde p_3,\tilde q_4,\ldots,\tilde q_{2n},\tilde p_{2n+1})  \nonumber \\
&& =\sum\limits_{k=1}^{n+1}\tilde p_{2k-1}q_k+F(q_1,\tilde q_2,\ldots,q_n,\tilde q_{2n},q_{n+1},\ldots,q_{2n+1}).    \label {w39}
\end{eqnarray}
Summarizing, we have found a modified Lagrangian and Hamiltonian formulations of higher-derivative theories. 
They are equivalent to the Ostrogradski 
formalism in the sense that on the Hamiltonian level they are related to the latter by a canonical transformation.
However, the advantage of the approach presented 
is that the Legendre transformation can be performed in a straightforward way. 
\vspace{12pt}
\par
{\large\bf Acknowledgement}
Thanks are due to Prof. P. Kosi\'nski for interesting discussions.
We are greatful to the unknown referee for very useful remarks.
 

\newpage
\begin{thebibliography}{99}
\addcontentsline{toc}{chapter}{Bibliografia}
\bibitem{1b}\textsc{W. Thiring,} Phys. Rev.  {\bfseries 77} (1950), 570.
\bibitem{2b}\textsc{A. Pais, G.E. Uhlenbeck,} Phys. Rev.  {\bfseries 79} (1950), 145.
\bibitem{3b}\textsc{K.S. Stelle,}  Phys. Rev.  {\bfseries D16} (1977), 953.
\bibitem{4b}\textsc{E.S. Fradkin, A.A. Tseytlin,} Nucl. Phys. {\bfseries B201} (1982), 469.
\bibitem{19b}\textsc{M.S. Plyushchay,} Mod. Phys. Lett. {\bfseries A4} (1989), 837 ; Mod. Phys. Lett. {\bfseries A3} (1988), 1299 ; Phys. Lett.{\bfseries B243} (1990), 383; \textsc{Yu.A. Kuznetsov, M. S. Plyushchay,} Nucl. Phys. {\bfseries B389} (1993), 181
\bibitem{5b}\textsc{M. Ostrogradski,}  Mem. Acad. St. Petersburg {\bfseries 4} (1850), 385.
\bibitem{6b}\textsc{J.M. Pons,} Lett. Math. Phys.  {\bfseries 17} (1989), 181.
\bibitem{7b}\textsc{T. Govaerts, M.S. Rashid,}  hep-th/9403009.
\bibitem{8b}\textsc{T. Nakamura, S. Hamamoto,} Prog. Theor. Phys.  {\bfseries 95} (1996), 409.
\bibitem{9b}\textsc{M. Henneaux, C. Teitelboim,}\emph{Quantization of gauge systems}, Princeton University Press 1992.
\bibitem{10b}\textsc{A. De Felice, S. Tsujikawa,} arXiv: 1002.4928.
\bibitem{11b}\textsc{N. Derulle, Y. Sendouda, A. Youssef,} Phys. Rev. {\bfseries D80} (2009), 084032.
\bibitem{12b}\textsc{H.J.  Schmidt,} Phys. Rev. {\bfseries D49} (1994), 6345.
\bibitem{13b}\textsc{H.J.  Schmidt,}  arXiv: gr-qc/9501019.
\bibitem{14b}\textsc{T.-C. Cheng, P.-M. Ho, M.-C Yeh,} \emph{} Phys. Rev. {\bfseries D65} (2002), 085015.
\bibitem{15b}\textsc{S. Hawking, T. Hertog,}  Phys. Rev. {\bfseries D65} (2002), 103515.
\bibitem{16b}\textsc{C. Battle, J. Gomis, J.M. Pons, N. Roman-Roy,} Journ. Phys.   {\bfseries A21} (1988), 2693.
\bibitem{18r}\textsc{P.~Dunin--Barkowski, A.~Slepsov} arXiv:~0801.4293
\bibitem{19r}\textsc{A.~Mironov, A.~Morozov} Theor.~Math.~Phys {\bfseries 156} (2008), 1209.\\
\textsc{A.~Mironov, A.~Morozov} Int.~J.~Mod.~Phys {\bfseries A23} (2008), 4686.\\
\textsc{D.~Galakhov} JETP Letters {\bfseries 87} (2008), 452.
\bibitem{17b}\textsc{V.V. Nesterenko,} Journ. Phys. {\bfseries A22} (1989), 1673.
\bibitem{18b}\textsc{K. Andrzejewski, J. Gonera, P. Ma\'slanka,} arXiv: 0710.2976.
\end{thebibliography}
\end{document}